\documentclass[pdftex,onecolumn,epjc3]{svjour3}	
\usepackage[a4paper]{geometry}				
\RequirePackage[T1]{fontenc}

\smartqed  

\RequirePackage{graphicx}
\RequirePackage{mathptmx}      
\RequirePackage{flushend}
\RequirePackage[numbers,sort&compress]{natbib}
\RequirePackage[colorlinks,citecolor=blue,urlcolor=blue,linkcolor=blue]{hyperref}

\usepackage{units}
\usepackage{color}
\usepackage{multirow}
\usepackage{amsmath}
\usepackage{amssymb}
\usepackage{verbatim}
\usepackage{subfig}

\journalname{Eur. Phys. J. C}

\begin{document}

\title{The prototype detection unit of the KM3NeT detector\thanksref{t1}}

\subtitle{KM3NeT Collaboration}

\author{
S.~Adri\'{a}n-Mart\'{i}nez\thanksref{UPV} \and
M.~Ageron\thanksref{CPPM} \and
F.~Aharonian\thanksref{DIAS} \and
S.~Aiello\thanksref{I-CAT} \and
A.~Albert\thanksref{GRPHE} \and
F.~Ameli\thanksref{I-ROM} \and
E.G.~Anassontzis\thanksref{U-ATH} \and
G.C.~Androulakis\thanksref{NCSR} \and
M.~Anghinolfi\thanksref{I-GEN} \and
G.~Anton\thanksref{ECAP}
% ...
}

\institute{
\label{UPV}{Instituto de Investigaci\'{o}n para la Gesti\'{o}n Integrada de las Zonas Costeras,Universitat Polit\`{e}cnica de Val\`{e}ncia,~Gandia,~Spain}
\and
\label{CPPM}Aix Marseille Universit\'{e}, CNRS/IN2P3, CPPM UMR 7346, 13288, Marseille, France
\and
\label{DIAS}DIAS, Dublin, Ireland
\and
\label{I-CAT}INFN, Sezione di Catania, Catania, Italy
\and
\label{GRPHE}GRPHE, Universit\'{e} de Haute Alsace, IUT de Colmar, Colmar, France
\and
\label{I-ROM}INFN, Sezione di Roma, Roma, Italy
\and
\label{U-ATH}National and Kapodistrian University of Athens, Deparment of Physics, Athens, Greece
\and
\label{I-GEN}INFN, Sezione di Genova, Genova, Italy
\and
\label{ECAP}Erlangen Centre for Astroparticle Physics, Friedrich-Alexander-Universit{\"a}t Erlangen-N{\"u}rnberg,Erlangen, Germany
% ...
\and
\label{I-NAP}INFN, Sezione di Napoli, Napoli, Italy
\and
\label{NCSR}Institute of Nuclear Physics, NCSR "Demokritos", Athens, Greece
% ...
\and
\label{U-NAP}Universit\`{a} 'Federico II', Dipartimento di Fisica, Napoli, Italy
\and
\label{I-BAR}INFN, Sezione di Bari, Bari, Italy
% ...
\and
\label{I-PIS}INFN, Sezione di Pisa, Pisa, Italy 
\and
\label{U-PIS}Universit{\`a} di Pisa, Dipertimento di Fisica , Pisa, Italy
\and
\label{I-LNS}INFN, Laboratori Nazionali del Sud, Catania, Italy
% ...
\and
\label{U-LEI}Leiden Institute of Physics, Leiden University, Leiden, The Netherlands
\and
\label{U-ROM}Universit{\`a} di Roma La Sapienza, Dipartimento di Fisica, Roma, Italy
\and
\label{U-SAL}Universit{\`a} di Salerno, Dipartimento di Fisica, Fisciano, Italy
\and
\label{U-UVA}Institute of Physics, University of Amsterdam, Amsterdam, The Netherlands
% ...
\and
\label{I-BOL}INFN, Sezione di Bologna, Bologna, Italy
\and
\label{U-BOL}Universit\`a di Bologna, Dipartimento di Fisica e Astronomia, Bologna, Italy
\and
\label{U-PSUD}Universit\'{e} Paris-Sud , 91405 Orsay Cedex, France
\and
\label{U-THS}Aristotle University Thessaloniki, Thessaloniki, Greece
\and
\label{U-WZB}University W{\"u}rzburg, W{\"u}rzburg, Germany
\and
\label{I-LNF}INFN, INFN, Laboratori Nazionali di Frascati, Frascati, Italy
% ...
\and
\label{U-TUB}Eberhard Karls Universit{\"a}t T{\"u}bingen, T{\"u}bingen, Germany
\and
\label{U-UU}Utrecht University, Utrecht, The Netherlands
\and
\label{U-BAM}Dr. Remeis Sternwarte, Friedrich-Alexander-Universit{\"a}t Erlangen-N{\"u}rnberg, Bamberg, Germany
\and
\label{U-CAT}Universit\`{a} di Catania, Dipartimento di Fisica ed Astronomia, Catania, Italy
\and
\label{U-CYP}University of Cyprus, Physics Department, Nicosia, Cyprus
\and
\label{U-AEG}University of Aegean, Athens, Greece
% ...
\and
\label{U-ABD}Oceanlab, University of Aberdeen, Aberdeen, United Kingdom
\and
\label{U-GEN}Universit\`{a} di Genova, Dipartimento di Fisica, Genova, Italy
% ...
\and
\label{U-SHF}University of Sheffield, Department of Physics and Astronomy, Sheffield, United Kingdom
% ...
}

\thankstext{ACCNL}{Also at Accademia Navale di Livorno, Livorno, Italy}
\thankstext{corr1}{Corresponding author: biagi@bo.infn.it}
\thankstext{corr2}{Corresponding author: creusot@apc.in2p3.fr}
\thankstext{corr3}{Corresponding author: dosamt@nikhef.nl}

\thankstext[$\star$]{t1}{The research leading to these results has received funding from the European Community Sixth Framework Programme under Contract 011937 and the Seventh Framework Programme under Grant Agreement 212525.}

\date{Received: date / Accepted: date}

\maketitle

\begin{abstract}
A prototype detection unit of the KM3NeT deep-sea  neutrino telescope  has been installed at $\unit[3500]{m}$ depth $\unit[80]{km}$ offshore the Italian coast.  
KM3NeT in its final configuration will contain several hundreds  of detection units.  
Each detection unit is  a  mechanical structure anchored to the sea floor, held vertical by a submerged buoy and supporting optical modules for the detection of Cherenkov light emitted by charged secondary particles emerging from neutrino interactions.  
This prototype string   implements three  optical modules with 31 photomultiplier tubes  each.  
These optical modules were developed by the KM3NeT Collaboration 
to enhance the detection capability of neutrino interactions.  
The prototype detection unit was operated since its deployment in May 2014  until its decommissioning in July  2015. 
Reconstruction of the particle trajectories from the data requires a nanosecond accuracy in the time calibration.  
A procedure for relative time calibration of  the   photomultiplier tubes  contained in each  optical module is described.  
This procedure  is based on the measured coincidences produced in the sea by the $^{40}$K  background light and  can easily be  expanded to  a detector with several thousands of optical modules.  
The time offsets between the different optical modules are obtained using LED nanobeacons mounted inside them. 
A  set  of data corresponding to 600 hours  of livetime was analysed.  
The results show  good agreement with   Monte Carlo simulations of the expected optical background and the signal from atmospheric muons. 
An almost  background-free sample of muons was selected by filtering the time correlated signals on all the three optical modules.  
The zenith angle of the selected  muons was reconstructed with  a precision of  about  $3^{\circ}$. 
\end{abstract}



\section{Introduction} 
\label{s:intro}

Following the scientific results obtained with the ANTARES  \cite{antares} neutrino telescope and the experience from  the NEMO  \cite{nemo} and NESTOR  \cite{nestor}  pilot projects, the KM3NeT Collaboration has started the construction of the  next generation deep-sea neutrino telescope   in the Mediterranean Sea \cite{km3net}.  
The main objectives of KM3NeT are the discovery and subsequent observation  of high-energy neutrino sources in the Universe and the determination of the neutrino mass hierarchy.
Neutrinos can  interact with matter inside or in the vicinity of the detector producing secondary particles that can be detected through the Cherenkov light that they produce.     
Due to the long range in water, the conventional  detection channel is given by muons produced in  charged current interactions of muon neutrinos.  
Furthermore, KM3NeT will have significant  sensitivity to all  the neutrino   interactions  \cite{maarten, luigi_icrc, paolo_icrc}.

The basic detection element of the neutrino telescope is the digital optical module (DOM), a 17-inch   pressure resistant  glass sphere containing 31 3-inch photomultiplier tubes (PMTs),   a number of calibration devices and the  read-out electronics.  
The multi-PMT design provides a large photocathode area ($\approx\unit[1400]{cm^2}$ per DOM \cite{expansion_cone}),  good separation between single-photon and multiple-photon hits and information on the photon direction.  

A group of 18 DOMs distributed in space along two thin ropes constitutes the essential part of a detection unit (DU).   
The bottom of the DU is anchored to the sea floor and is kept close to vertical  by a submerged buoy.  
The DUs are connected to shore via a sea-bottom network of electro-optical cables and junction boxes.  
Data collected by the PMTs are   digitised in the DOMs and sent to shore, where they are filtered by appropriate triggering algorithms. 
Accurate measurements of the light arrival times and charges and precise real-time knowledge of the positions and orientations of the PMTs are required for the accurate reconstruction of the direction of the secondary particles.

A prototype DOM (Pre Production Model DOM, PPM-DOM) was deployed in April 2013 at the ANTARES site, $\unit[40]{km}$  offshore the French coast close to Toulon, attached to one ANTARES line  \cite{ppm-dom}. 
This   project has validated the DOM concept and technology demonstrating the capability of a single DOM to identify muons using time coincidences between PMTs inside one DOM.

In May 2014, a prototype detection unit (Pre Production Model DU, PPM-DU) with   3 DOMs was installed $\unit[80]{km}$ offshore the Sicilian coast.  
This prototype, unlike the PPM-DOM,  implements the  mechanical structure, the electro-optical connections and the data transmission system developed for the final  DU design.
In this configuration  for the first time  simultaneous data taking of several DOMs was proved in the deep sea.
Through the study of correlated signals in different DOMs generated from LED nanobeacons and from atmospheric muons, a synchronisation at a nanosecond level between DOMs was obtained.

In this paper  the main results obtained with this   project are presented, using data collected   between May 2014 and January 2015. 
In Section 2, an overview of the detector elements is given; the procedure of time calibration is described in Section 3; an evaluation of the optical background at the deployment site of the prototype  is provided in Section 4; Monte Carlo (MC) simulation are presented in Section 5; the capability to identify the signals from  muons and reconstruct their directions using inter-DOM coincidences is presented in Section 6.

\section{Detector} 
\label{ss:detector}

\begin{figure}
\begin{center}
\includegraphics[width=10cm]{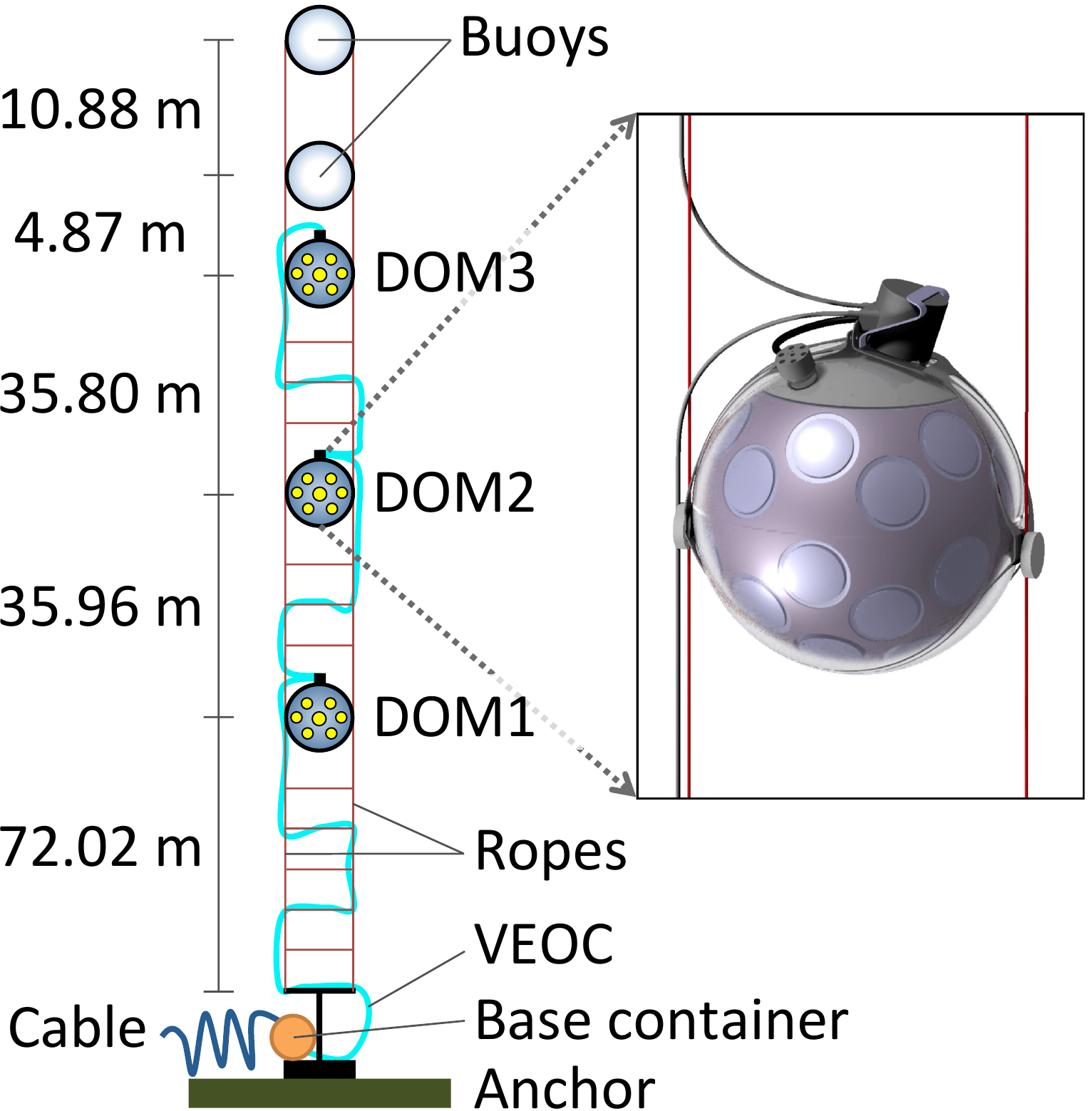}
\caption{Schematic of the PPM-DU (not to scale). Adjacent   DOMs are vertically spaced by $\sim\unit[36]{m}$.   
Two empty glass spheres serve as buoys; the vertical  electro-optical cable (VEOC) connects the DOMs with the base container, which is equipped with a $\unit[100]{m}$ cable  for connection to the submarine infrastructure and thus to the shore station.
Inset: the DOMs are attached to two Dyneema$^\circledR$ ropes; the structure is free to move following  underwater sea currents. }
\label{fig:ppmdu}
\end{center}
\end{figure}

The  PPM-DU  was deployed in May 2014 at $\unit[3457]{m}$ sea depth in a location $\sim\unit[80]{km}$ east of the Sicilian coast at Capo Passero (latitude $36^\circ\,17\text{'}\,50\text{''}\,\text{N}$, longitude $15^\circ\,58\text{'}\,45\text{''}\,\text{E}$).
The $\unit[160]{m}$ long PPM-DU comprises  three DOMs with a vertical separation of $\sim\unit[36]{m}$.  
It is anchored on the sea bottom  and is kept taut  by the buoyancy of the DOMs and two top flotation spheres  (Fig.~\ref{fig:ppmdu}).
The PPM-DU base is connected via an electro-optical cable to the cable termination frame   of the main electro-optical cable of the sea-bed network.
This $\unit[100]{km}$ long   cable bridges the  distance between the deep sea infrastructure and the shore station for power distribution and data transmission.

\subsection{The DOMs} \label{sec:doms}

The three  DOMs of the PPM-DU contain  two different types of PMTs with similar performance  but slightly different dimensions. 
The PMT model installed in DOM~1 and  DOM~2 is  the D783KFLA produced by ETEL \cite{vlvnt1}, while  DOM~3 contains the R12199-02  Hamamatsu PMTs \cite{vlvnt2}. 
The nominal diameter of the cathode area for the ETEL PMT is $\unit[72]{mm}$ and for the Hamamatsu PMT $\unit[76]{mm}$. 
Inside the DOM, the PMTs are surrounded by reflector rings at an angle of 45 degrees with respect to the PMT axis, $\unit[16]{mm}$ in width for ETEL and $\unit[17]{mm}$ for the Hamamatsu PMTs \cite{expansion_cone}.
The PMTs are operated at a gain of $3\times10^6$ with an intrinsic dark count rate in the range $\unit[600\text{--}1500]{Hz}$, as measured in the laboratory at  room temperature with a threshold of 0.3~photoelectrons (p.e.).  
Two PMTs channels,  one in DOM~2 and one in DOM~3, were not functional.

Each DOM contains the electronics boards and a power conversion board to readout, control and power all the PMTs, as well as sensors for acoustic  measurements and the monitoring of environmental conditions \cite{ppmdu}. 
A LED nanobeacon \cite{nanobeacon} with a wavelength of $\unit[470]{nm}$ is installed in the upper part of each DOM, pointing upwards. 
The intensity and  frequency of flashing of the LED nanobeacon signals  are controlled  from shore.

\subsection{String design} \label{sec:string}

A schematic  of the PPM-DU is shown in Fig.~\ref{fig:ppmdu}.  
The three DOMs are attached to two thin parallel Dyneema$^\circledR$ ropes  (inset of  Fig.~\ref{fig:ppmdu}). 
A vertical electro-optical cable (VEOC),  an oil filled plastic tube containing copper wires and optical fibres for power and data transmission,   is attached to the ropes and provides breakouts to each DOM.
Additional empty  spheres at the top of the string increase the buoyancy for  keeping  the string  close to  vertical. 
The base anchors the string to  the sea bottom and houses a power converter and  fibre-optic components.

The PPM-DU was mounted on a launcher vehicle and deployed on the sea bottom.  
The  launcher vehicle  is a   spherical structure with a diameter of  $\sim\unit[2]{m}$      designed to accommodate and deploy a full-size DU  of KM3NeT \cite{LOM}.  
Once on the sea bottom, an acoustic release initiates the unfurling of the string,  and the launcher vehicle  floats to the surface to be recovered.  
Connection of the string to the under-sea infrastructure is performed by means of a Remotely  Operated  Vehicle (ROV).

\subsection{Data acquisition  } \label{sec:daq}

The detector readout follows the  ``all-data-to-shore'' approach  \cite{all_data_to_shore}.
The readout electronics board (Central Logic Board, CLB \cite{clb}) inside the DOM provides for the  data acquisition and communication with the shore station. 
Each DOM is an IP node in an Ethernet network.   
The information recorded from a PMT consists of the start time and the time over threshold (ToT).
The start time is defined as the time at which the pulse passes beyond a $\unit[0.3]{p.e.}$ threshold and the ToT is the time   the pulse remains above this threshold.
The ToT signals from the PMTs pass to the CLB where they are time stamped and arranged in timeslices   of $2^{24} \cdot \unit[8]{ns} \approx \unit[134]{ms}$.
Data are transferred to shore via the optical fibre network.  
Onshore, the physics events are filtered from the background by an online trigger algorithm and stored on disk. 
The  Level-1 trigger is defined as two hits in a DOM in separate PMTs, with a time difference smaller than $\unit[25]{ns}$ (L1 hit).
A physics event is  triggered  grouping all L1 hits with a time difference smaller than $\unit[330]{ns}$  which is consistent with signals from  particles passing in the vicinity of the detector.   
In addition to the physics events, summary data containing all the singles PMT rates are recorded and stored on disk.  

The first four  months of data taking were used to test the system and optimise the operation which results in an average livetime of  $\sim\unit[18]{hours/day}$.      
The dead-time is due to the necessary periodic initialisations of the CLB and the transfer of data from the PC acquiring the data to another location for further filtering and distribution  which cannot happen simultaneously with data taking in this prototype.
Both issues are resolved in the design of the full-scale KM3NeT detector.

\section{Time calibration}
\label{s:calib}

A time calibration  at a nanosecond  level is necessary to achieve the envisaged angular resolution for a neutrino telescope.
For this, the following time offsets have to be determined:
\begin{itemize}
\item Intra-DOM time offsets (between PMTs in the same DOM) that   primarily depend on the PMT transit  time; 
\item Inter-DOM time offsets (between DOMs)  that   primarily depend on the cable lengths.  
\end{itemize}
For the intra-DOM offsets  signals from $^{40}$K decays are exploited, while for the inter-DOM time offsets  calibration runs with the LED nanobeacons are used.

\subsection{Intra-DOM calibration} \label{ss:k40}

\begin{figure}
\centering
{\includegraphics[width=8.5cm]{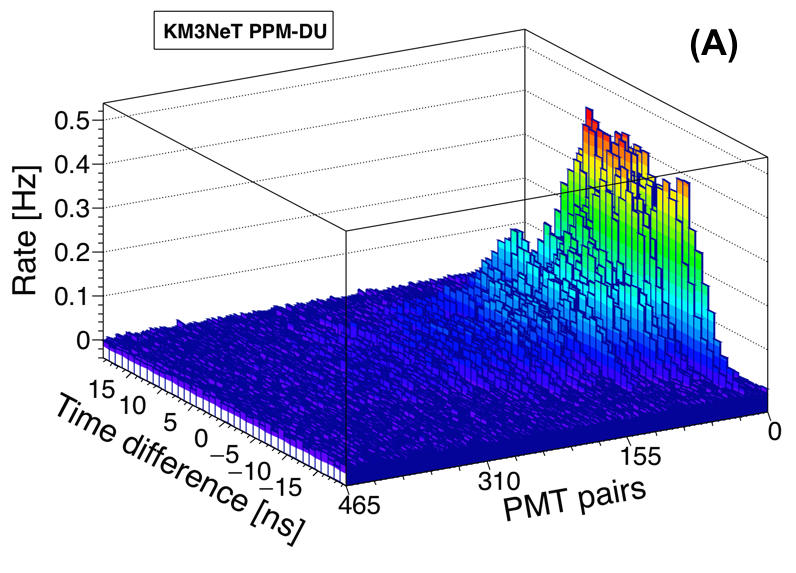}} \\ 
{\includegraphics[width=8.5cm]{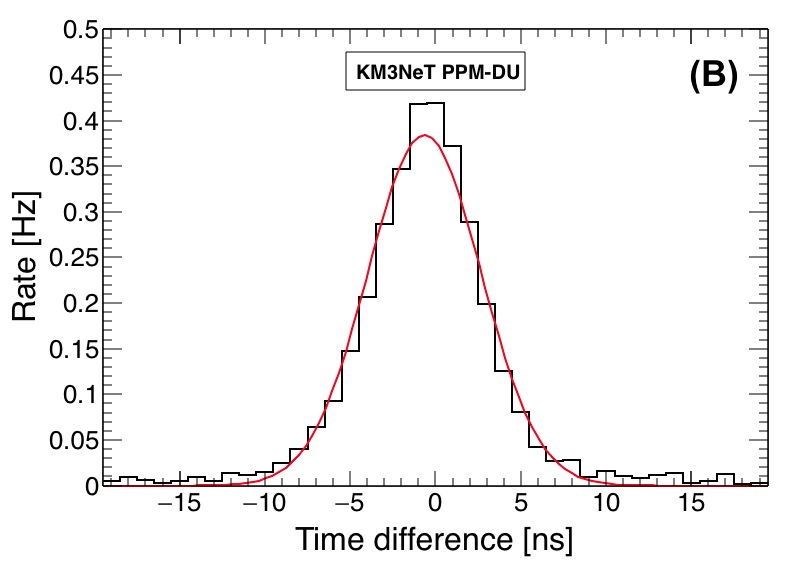}}
\caption{(a)  Distribution of  time differences between the hit  times for all   PMT pairs in  DOM~1 for one physics run. 
The PMT pairs are ordered according to the angular distance between PMTs.
(b) Distribution of time differences between the hit  times of two adjacent PMTs in  DOM~1 for one physics run. The Gaussian function,  represented by the red line, is the result of the simultaneous fit (see text). The baseline  due to combinatorial background has been subtracted from the data.}
\label{f:deltaT}
\end{figure}

Radioactive decays of $^{40}$K present in sea water typically produce up to 150 Cherenkov photons per decay \cite{zaborov_antares}.
These decays are the main source of the singles rates observed in the PMTs.
A single decay occurring in the vicinity of the DOM has a chance to produce a genuine coincidence between signals of different PMTs, which can be exploited for time calibration of the DOM.
The procedure to obtain the time offsets between PMTs in one DOM  from the signal time coincidences follows the approach described in Ref. \cite{ppm-dom}.  
The distributions of time differences between signals detected in different PMTs in the same DOM are studied as a function of the angular separation of the  PMTs involved.  
The distribution of hit time differences between all possible combinations of PMT pairs are assumed to follow a Gaussian shape.  
For each DOM with $N$=31 PMTs, a total of $N (N-1)/2$ distributions are produced and shown in Fig.~\ref{f:deltaT}a for DOM~1.   
In the figure,  the numbers of   PMT pairs are ordered according  to their angular separation.   
The  correlation peak decreases as the angular separation increases due to the limited field of view of each PMT.
An example of time differences between two adjacent PMTs of DOM~1 is given in Fig.~\ref{f:deltaT}b.    
To obtain the rate of coincidences shown in the figure, the flat  combinatorial background due to uncorrelated hits on the two PMTs has been subtracted.
These distributions are well fitted by a Gaussian function.
The mean values, heights and widths of the Gaussian peaks are related to the time offsets, detection efficiencies and intrinsic time-spreads of all the PMTs. 
Typically, a FWHM of $\unit[7\text{--}10]{ns}$ is found for all different PMT pairs,   
mostly reflecting the intrinsic PMT transit time spread of up to $\unit[5]{ns}$   at  FWHM.

\begin{figure}
\centering
\includegraphics[width=8.5cm]{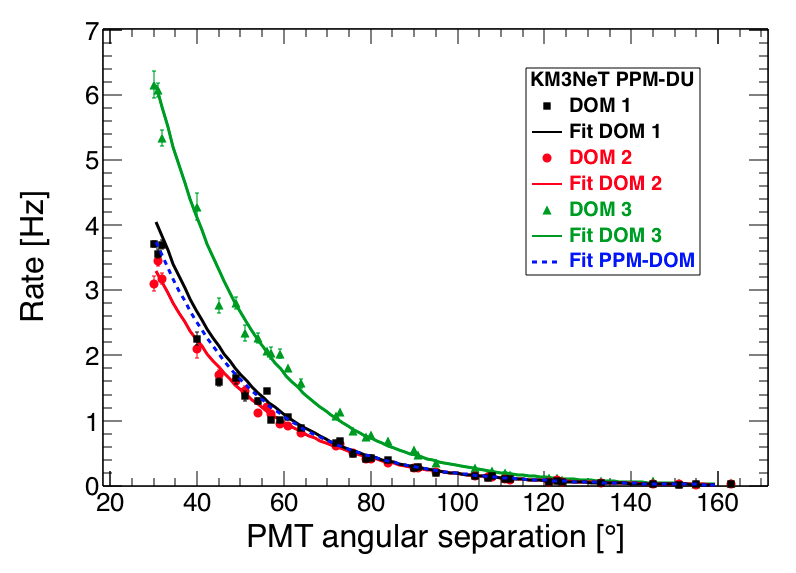}
\caption{Rate of twofold coincidences  as a function of the angular separation between two PMTs, for the DOMs of the PPM-DU. 
The curves are the result of fitting an exponential function to the data.}
\label{f:2foldrateVsAngle}
\end{figure}

The twofold coincidence rate is  shown as a function of the angular separation between pairs of PMTs in Fig.~\ref{f:2foldrateVsAngle} for the three  DOMs of the PPM-DU.
The angular dependence for all PMT pairs can be fitted to an exponential function as shown in Fig.~\ref{f:2foldrateVsAngle}.
The scattering  of  data points is partially due to the slightly different PMT efficiencies.  
The significantly higher rate of DOM 3 can be explained as the Hamamatsu PMTs have   larger photo-cathode area and larger reflector rings.  
Data from the PPM-DOM have been analysed  using the same procedure adopted in this work and shown in Fig.~\ref{f:2foldrateVsAngle} for comparison.

\begin{figure}
\centering
{\includegraphics[width=8.5cm]{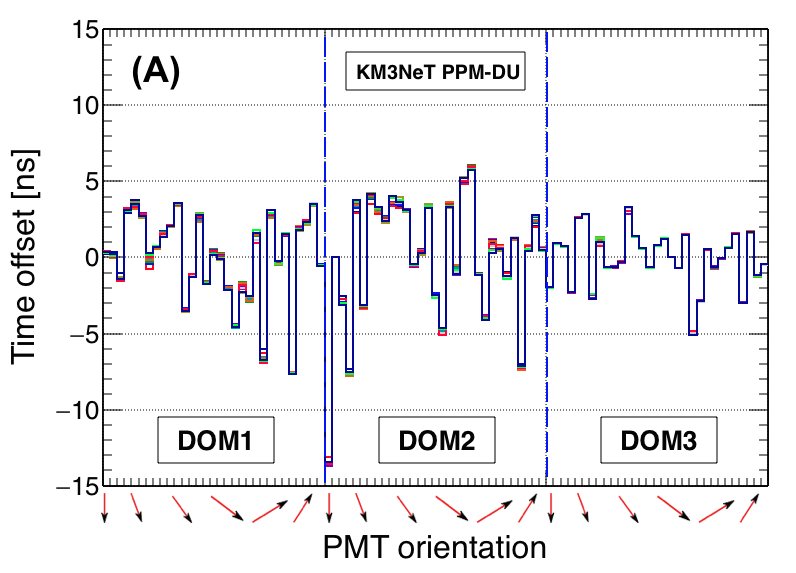}} \\ 
{\includegraphics[width=8.5cm]{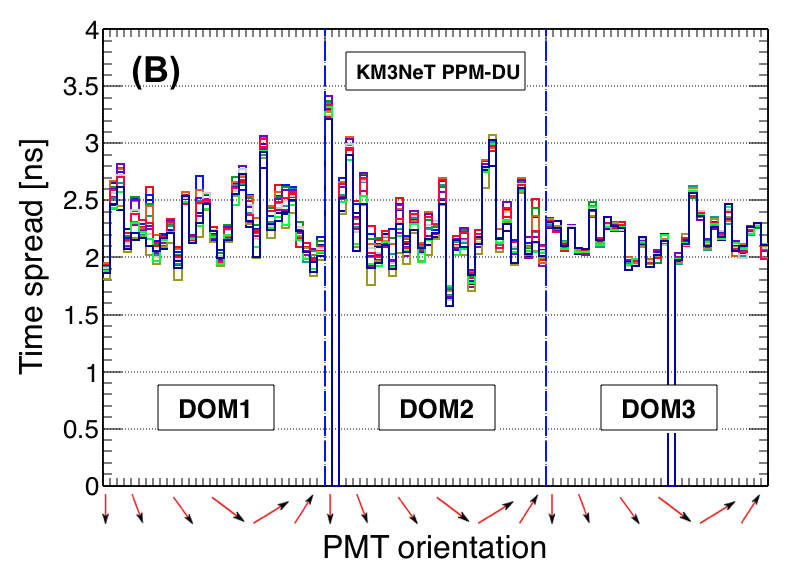}} \\ 
{\includegraphics[width=8.5cm]{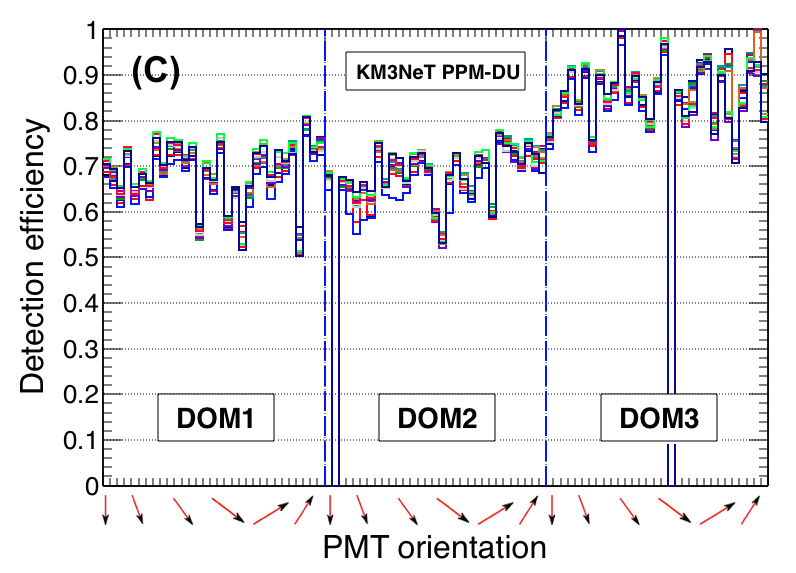}}
\caption{Results of the intra-DOM calibration procedure.  
Different colours refer to  13   data runs of 30 minutes each.  
(a) Relative time offsets of the  PMTs inside a DOM.   The average of time offsets inside a DOM is set to zero. 
(b) Intrinsic time-spreads of the PMTs  expressed as the Gaussian sigma.   
(c) Relative detection  efficiencies  of the PMTs. All   detection efficiencies are normalised to the overall  highest one, that is set to unity.  The PMT  photo-cathode area  enters the detection efficiency, resulting in   larger values for  DOM~3.
The respective values of the two defective PMTs have been set to zero in all three plots.}
\label{f:GlobalFit}
\end{figure}

Assuming the  exponential angular dependence of the coincidence rates due to $^{40}$K reported for each DOM in Fig.~\ref{f:2foldrateVsAngle}, a $\chi^2$ minimisation procedure is applied  to obtain simultaneously the relative time offsets, the detection efficiencies and the intrinsic time-spreads of all  PMTs in a DOM. 
The results of the fitting procedure are shown in Fig.~\ref{f:GlobalFit} for 13 different physics data runs randomly selected. 
It is observed that the relative time offsets and  PMT detection efficiencies are stable over time scales of months. 
The resulting relative time offsets are found to be mostly  less than $\unit[10]{ns}$.  
The relative time offsets obtained with this method  are stable  in time  within $\unit[0.5]{ns}$. 
The relative detection  efficiencies of the same PMT type differ by less than 10\%  and are  stable in time  within 3\%.

\subsection{Inter-DOM calibration} \label{ss:beaconCalib}

\begin{figure}
\centering
{\includegraphics[width=8.5cm]{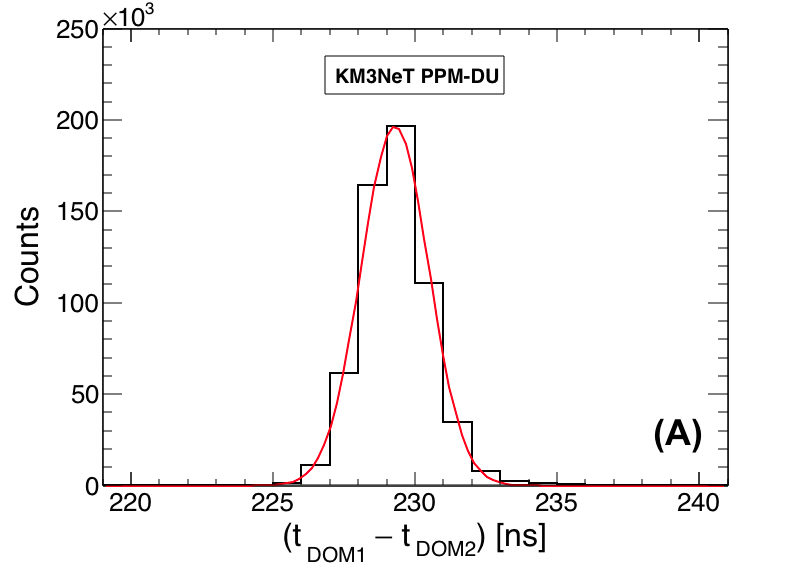}} \\ 
{\includegraphics[width=8.5cm]{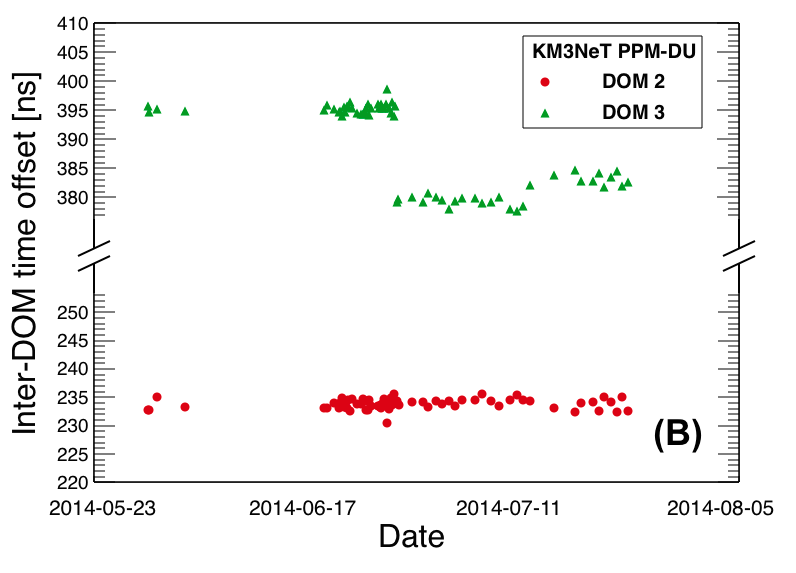}}
\caption{(a) Distribution of the time differences between hit detections on DOM~1 and DOM~2 when operating  the LED nanobeacon for one run. 
The distribution  is corrected for the expected light travel time.  
(b) Mean time offsets for DOM~2 and DOM~3 with respect to DOM~1 for different LED nanobeacon runs. }
\label{f:domDeltaT}
\end{figure}

LED nanobeacon runs are used to calculate the inter-DOM  time offsets.   
The time differences between pairs of DOMs  are  calibrated in   runs in which the LED nanobeacon of the lowest  DOM   is operated. 
The results are cross checked with results from data where the LED nanobeacon of the middle DOM is  operated and good consistency is found.
The distribution of  time differences of coincident hits on the DOM with the nanobeacon and the DOM to be calibrated are corrected for the travel time of light in the sea water.   
The distribution is then fitted with a Gaussian function as shown in Fig.~\ref{f:domDeltaT}a. 
For the travel time of  light, a fixed  distance between the nanobeacon and the hit PMT is used. 
In the calculation, the group velocity  of $\unit[470]{nm}$ light (nanobeacon wavelength) in  water is used.   
The histogram of  time differences between DOM~1 and 2 is shown in Fig.~\ref{f:domDeltaT}a.  

The resulting mean time offsets per run for DOM~2 and 3 with respect to DOM~1 are shown in   Fig.~\ref{f:domDeltaT}b.  
The changes in the time offset of DOM~3 are due to power cycles of the corresponding module in the shore station.
Shifts of this size are to be expected in the prototype as a full time synchronisation is not   implemented.  
No time offset shifts were observed within constantly powered periods. 
The values obtained with this procedure are stable to within a few nanoseconds  over a stable period of data taking.  
The calibration using nanobeacons was cross checked   with an alternative calibration procedure using  the signal from muons.
Agreement within $\unit[2]{ns}$ was found (see Sect. \ref{s:multiDOM}).

\section{Singles and multi-fold coincidence rates}
\label{s:analysis}

\begin{figure}
\centering
{\includegraphics[width=8.5cm]{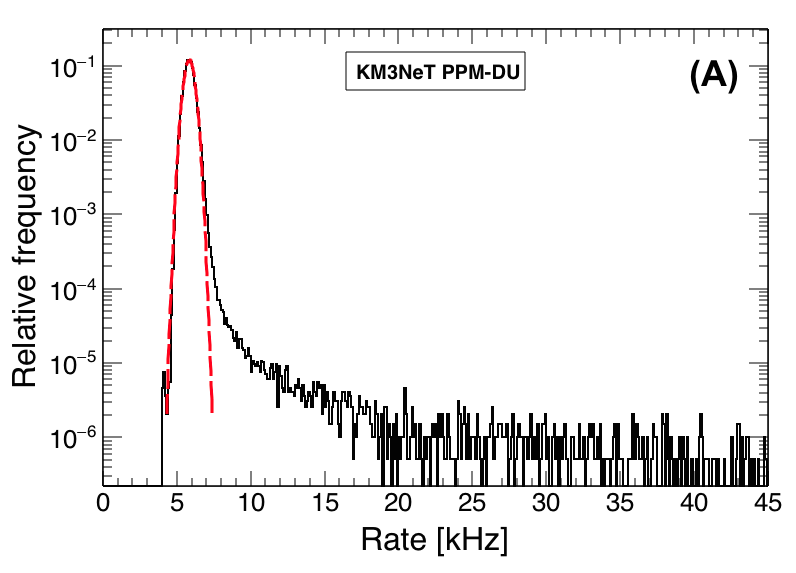}} \\ 
{\includegraphics[width=8.5cm]{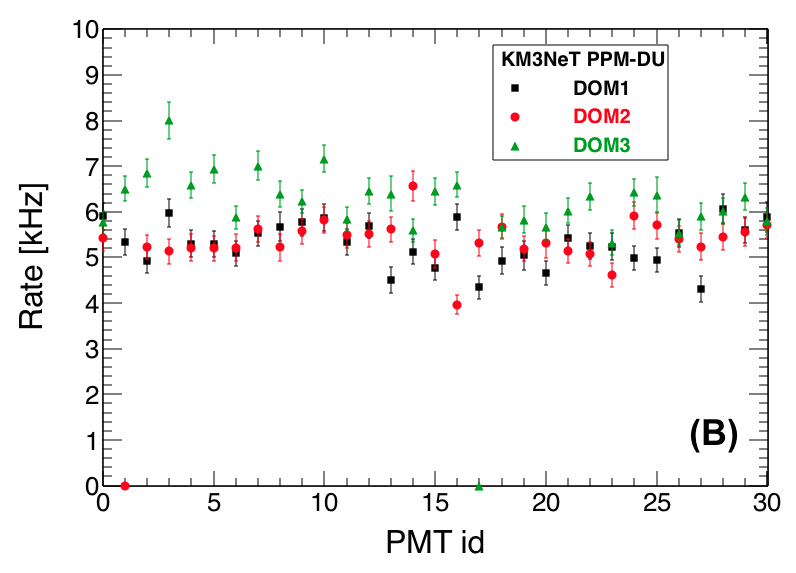}}
\caption{(a) Rate  distribution of PMT~16 of DOM~1 for the  whole data set, one entry per timeslice. 
(b) Mean value  of  the singles rates per PMT for the 3 DOMs.}
\label{f:singlerates}
\end{figure}

The two main contributions to the singles rates are the $^{40}$K decay and the bioluminescence activity.  
While the $^{40}$K decay is stable as a function of time and location in the detector,  the bioluminescence activity can fluctuate significantly in  time.  
The distribution of the average singles rate per timeslice of $\unit[134]{ms}$ for one PMT  of DOM~1 is shown in Fig.~\ref{f:singlerates}a.  
There are two easily identifiable contributions to the rate: a Gaussian distribution peaking at  $\sim\unit[5.9]{kHz}$ and a high frequency tail.  
The Gaussian peak is mainly due to  $^{40}$K decays. 
In Fig.~\ref{f:singlerates}b, the mean values of the Gaussian fit are plotted for all PMTs of the three DOMs.  
The horizontal axis refers to the PMT numbering scheme, with PMT 0 looking down, the next 6 PMTs being the ones in the lower ring, followed by those PMTs in subsequent rings.  
The   value of each data point corresponds to the mean of the Gaussian fit, and the error reflects the standard deviation of the fit.   
The average values for each DOM are given in Table~\ref{t:rates}.    
The singles rates are consistent with a fit of an exponential function to the distribution of time differences between consecutive hits, indicating that most of the singles rates is due to random background.

\begin{figure}
\centering
\includegraphics[width=8.5cm]{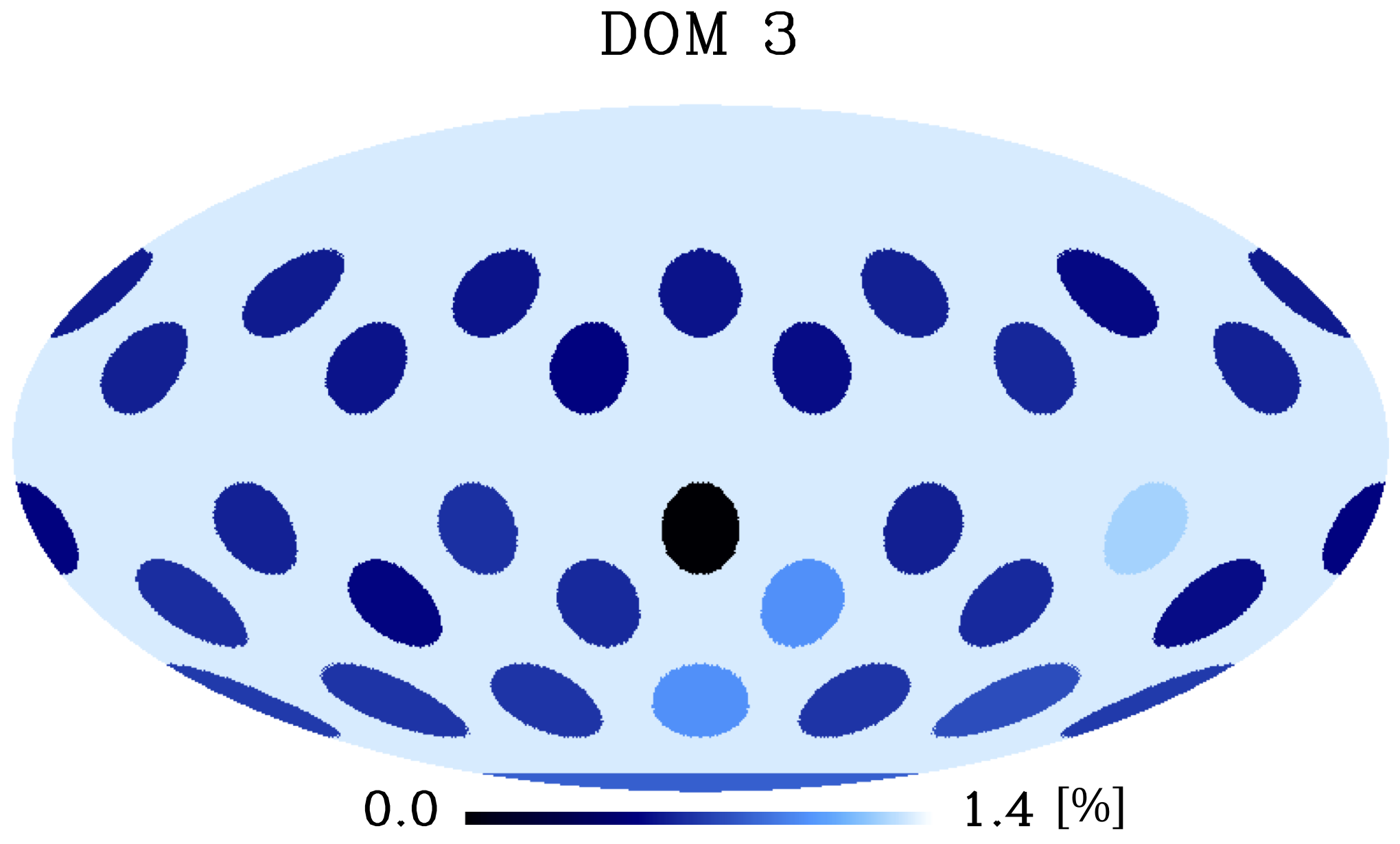} \\ 
\includegraphics[width=8.5cm]{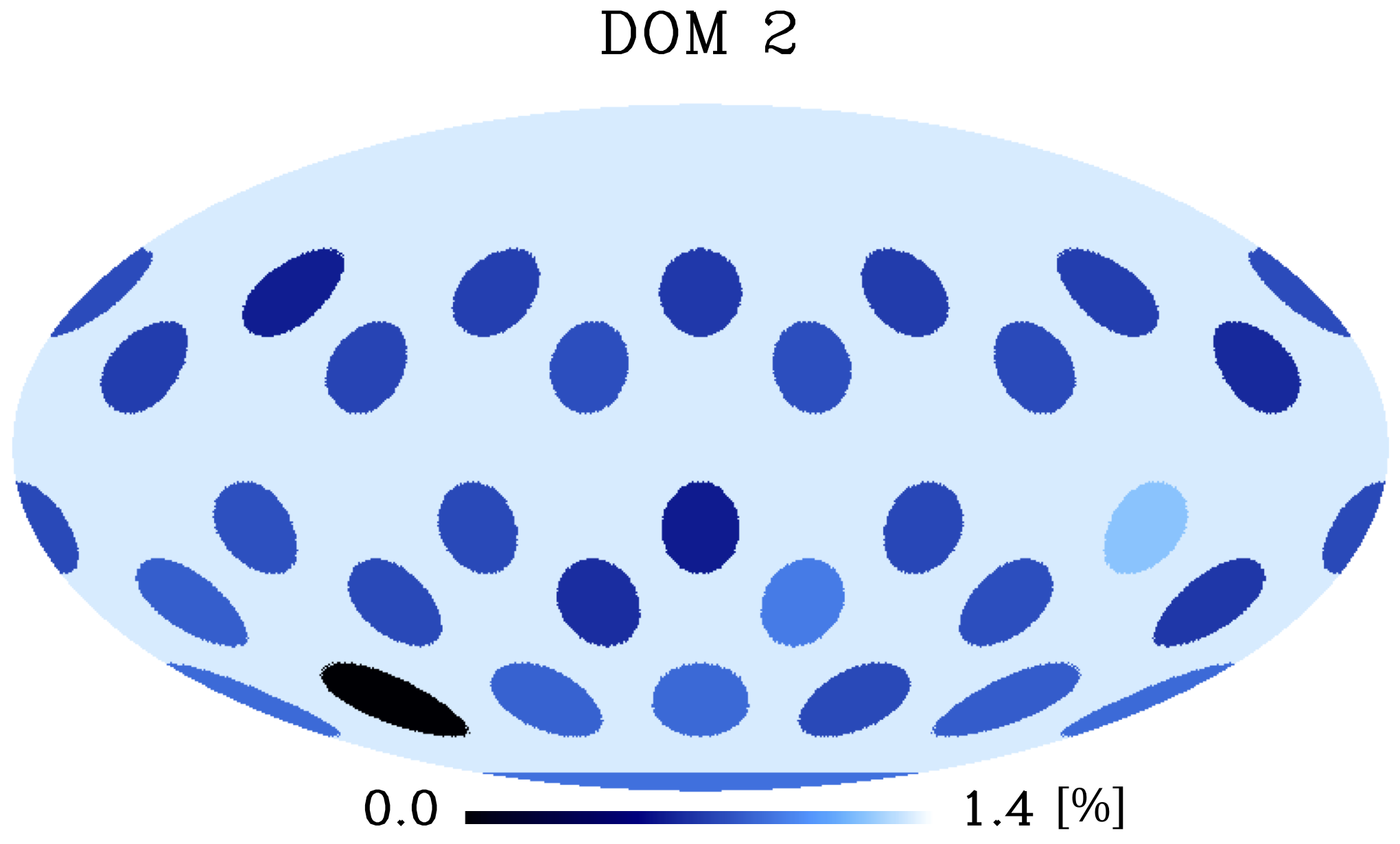} \\ 
\includegraphics[width=8.5cm]{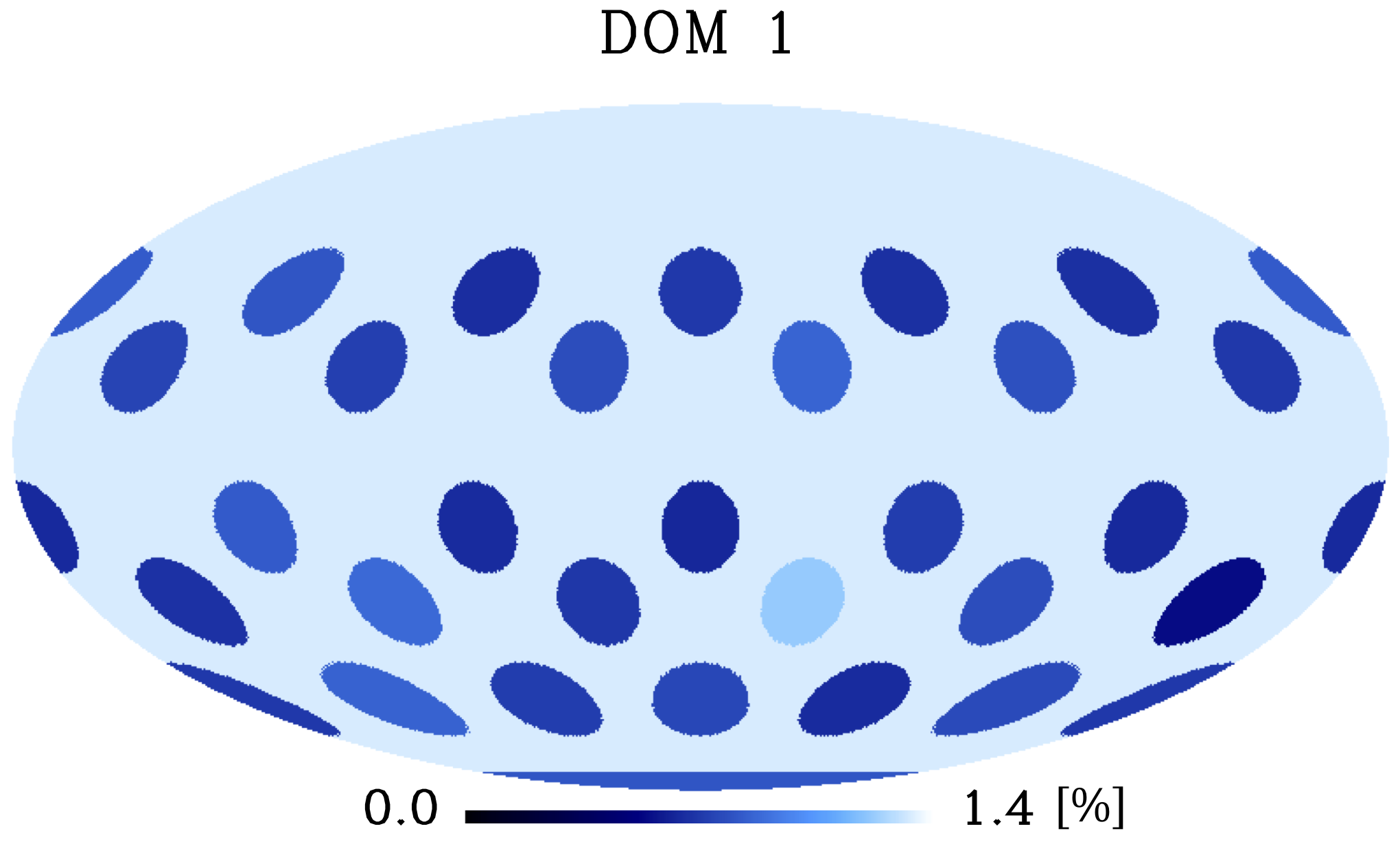}
\caption{Burst fraction percentage  of the 3 DOMs.  
The PMT positions are reported in the Aitoff azimuthal map projection. 
Black spots refer to PMTs that are not functional. 
}
\label{f:biolum1}
\end{figure}

The second contribution in the histogram of Fig.~\ref{f:singlerates}a is due to sporadic bioluminescence  background. 
It corresponds to timeslices with a high hit rate, typically three  or more  standard deviations above the mean of the Gaussian peak.    
These noisy timeslices are used to define the burst fraction as the ratio of the number of noisy slices to the total number of slices.   
The burst fraction for each DOM is shown in the three  plots of  Fig.~\ref{f:biolum1}.   
Contrary to what has been observed with the PPM-DOM, where the spatial distribution of the bioluminescence activity was attributed to the presence of the support structure of  the PPM-DOM \cite{ppm-dom}, there is no pattern in the spatial distribution of the bursts over the DOMs.    
The bioluminescence sporadic activity is thus homogeneous in the DOM vicinity.    

\begin{figure}
\centering
{\includegraphics[width=8.5cm]{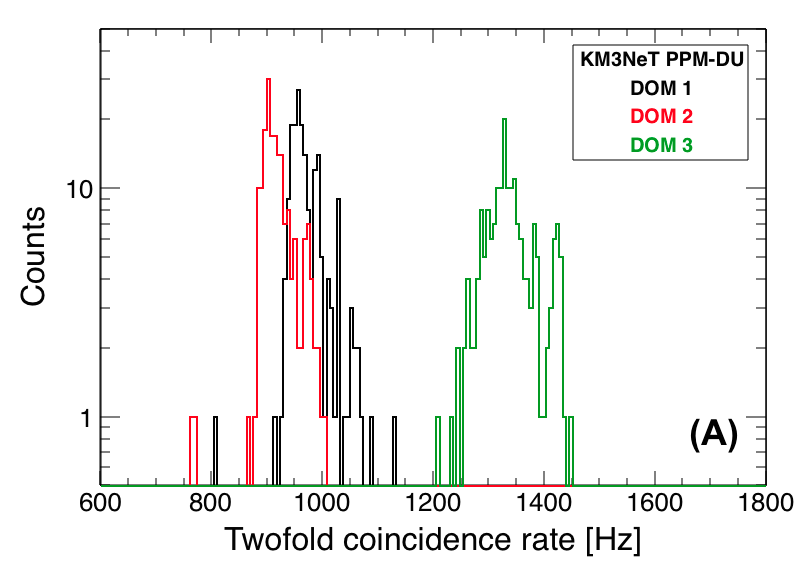}} \\ 
{\includegraphics[width=8.5cm]{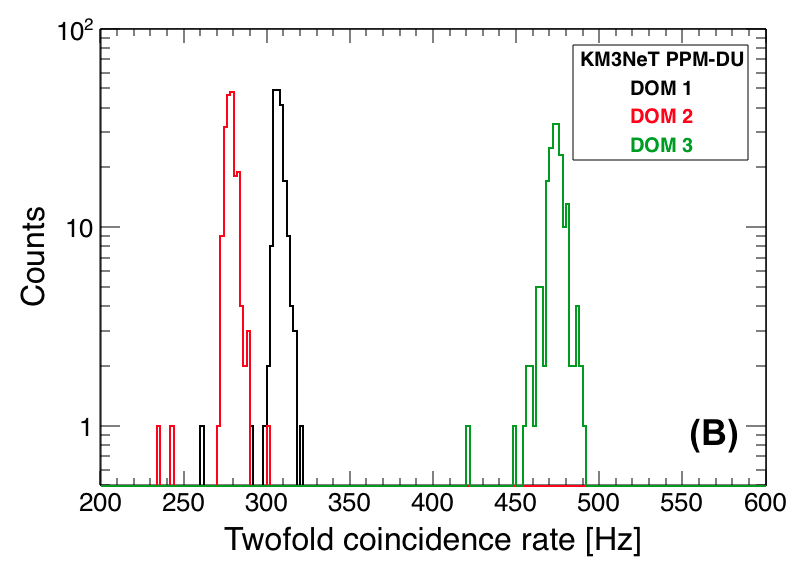}}
\caption{
Distributions of the rates of twofold coincidences (one entry per run) for the three DOMs, for a   coincidence time window of $\unit[25]{ns}$, (a) before and (b) after combinatorial background subtraction.}
\label{f:2fold}
\end{figure}

The distribution of the twofold coincidence rates (one entry per run) in a coincidence window of $\unit[25]{ns}$ is shown in Fig.~\ref{f:2fold} before (a) and after (b) combinatorial background subtraction. 
The rate of  coincidences due to combinatorial background is estimated through   Monte Carlo simulations assuming the PMT singles rates recorded in the summary data. 
Scrambled data  samples  are  simulated  randomising the hit time  within the timeslice and the rate of combinatorial background is obtained. 
The average value for each distribution  after background subtraction is given in Table~\ref{t:rates}. 
The rate was found to be stable over the observation time of seven  months  within a few percent. 
The combinatorial background did not show major variations. 
The differences between the DOMs are due to the  different  PMT efficiencies which enter here in square. 

Higher than twofold coincidences have also been studied.
The average values after combinatorial background subtraction for three- to sixfold and higher coincidences are given in Table~\ref{t:rates} for the three DOMs of the PPM-DU. 
The ratio between the DOM rates is expected to reflect the ratio between the DOM efficiencies to the power of the coincidence multiplicity. 
This is approximately the case up to a coincidence multiplicity of five. 
Above this value, the signal on PMTs  is not due to  processes (like  $^{40}$K decay) predominantly producing single photoelectrons and this relation does not hold anymore. 
It is worth  mentioning  that above a coincidence multiplicity of three  the combinatorial background becomes negligible.

\begin{table}
\centering
\begin{tabular}{c|c|c|c}
Coincidences & DOM 1 [Hz] & DOM 2 [Hz] & DOM 3 [Hz] \\
\hline
    Single & $(166\pm4)\cdot 10^3$ & $(162\pm12)\cdot 10^3$ & $(188\pm14)\cdot 10^3$\\
    2-fold & $307\pm5$ & $278\pm5$ & $473\pm7$\\
    3-fold & $23.1\pm0.5$ & $18.6\pm0.7$ & $44.1\pm0.9$\\
    4-fold & $2.03\pm0.07$ & $1.35\pm0.08$ & $4.89\pm0.19$\\
    5-fold & $0.17\pm0.02$ & $0.10\pm0.02$ & $0.53\pm0.04$\\
    6-fold & $0.018\pm0.005$ & $0.012\pm0.005$ & $0.057\pm0.011$\\
$>$ 6-fold & $0.017\pm0.006$ & $0.017\pm0.006$ & $0.030\pm0.011$
\end{tabular}
\caption{Mean coincidence rates for the three DOMs,  for a     coincidence time window of $\unit[25]{ns}$. The results are summed over the whole DOM. Note that in  DOM~2 and DOM~3 only 30 PMTs are involved in the data acquisition. 
}
\label{t:rates}
\end{table}

\section{Monte Carlo  simulation}
\label{s:mc}

A detailed Monte Carlo simulation of atmospheric muons has been performed, taking into account the relevant physics processes and the detector response.  
Cosmic rays entering the atmosphere produce extensive air showers containing high-energy muons.  
Although the sea water above the detector serves as a shield, many of these muons reach the detector, constituting the    source of  physics  signals for the PPM-DU.  
The simulation framework is based on the ANTARES software \cite{ANTsoftware}, modified to take into account the DOM properties.  
The simulation chain consists of the generation of atmospheric muons, their propagation in sea water, the generation and propagation of Cherenkov light, the $^{40}$K and bioluminescence background and the digitisation of the PMT signals.  
The simulation is based on a fixed detector geometry.   
The optical properties of the sea water and the PMT characteristics are taken into account in the simulation.   
The depth of the deployed string and the optical water properties measured at the   Capo Passero site have been used \cite{KM3NeT_TDR}.

Atmospheric muons were generated with the fast MUPAGE code \cite{MUPAGE}.    
This code provides a parameterisation of the underwater flux of atmospheric muons including multi-muon events (muon bundles)  based on a full Monte Carlo simulation of primary cosmic ray interactions and shower propagation in the atmosphere.   
Atmospheric muons were generated with energy $\text{E}_{\text{b}}>\unit[10]{GeV}$, where $\text{E}_{\text{b}}$ is the sum of the energies of the muons in the bundle.    
A sample statistically equivalent to a live time of 15.3 days  was generated.  

The generated muons were  tracked in  sea water with the code KM3 \cite{ANTsoftware}. 
This program uses tabulated results from full GEANT3.21 simulations of relativistic muons and electromagnetic cascades in sea water to generate the number of Cherenkov photons detected by the PMTs. The simulation takes into account the full wavelength dependence of Cherenkov light production, propagation, scattering and absorption in sea water, the response of the PMTs, including absorption in the glass and the optical gel, the PMT quantum efficiency, and the reduced effective area for photons arriving off-axis.
Light due to the background from  $^{40}$K  decays in sea water has been simulated by adding singles hit rate of $\unit[5.5]{kHz}$ per PMT and a time-correlated hit rate of $\unit[697]{Hz}$, $\unit[57]{Hz}$ and $\unit[7]{Hz}$ per DOM corresponding to two, three and four coincident hits in different PMTs in the same DOM, respectively.   
These parameters have been estimated via a detailed simulation based on GEANT4 \cite{GEANT}.  
An average dark current rate of $\unit[0.7]{kHz}$ per PMT has been also taken into account. 
The PMT detection efficiencies as estimated in Sect.~\ref{ss:k40} and shown in Fig.~\ref{f:GlobalFit}c have been used.

As mentioned in Sect.~\ref{ss:detector},  the start time and the ToT are recorded for each PMT hit.
This scheme is implemented in the detector simulation, with a smearing of the raw hit times 
that follows the measured PMT transit time response \cite{vlvnt1}.  
The ToT dependence on the number of photo-electrons and their time sequence for multiple hits on one PMT that cannot be resolved in time were also considered.
These corrections result in events containing complete and unbiased snapshots of all hits recorded during a time window around the atmospheric muon event.
The same trigger algorithm which was applied to the data was also applied to the simulated hits.
Triggered events containing only $^{40}$K hits  were also simulated.

\section{Muon reconstruction} 
\label{s:reconstruction}

As it has already been demonstrated by the PPM-DOM deployed at the KM3NeT French site \cite{ppm-dom}, even with a single DOM it is possible to reject the background and identify muon induced signals  by selecting  high multiplicity ($\geq 6$) coincident hits.  
In the case of the   PPM-DU, the correlated information from all 3 DOMs provides an extra handle for the identification of muons and allows for  a more precise reconstruction of the direction of the associated particles.  

\subsection{Muon signature in multifold coincidences}

\begin{figure}
\centering
\includegraphics[width=8.5cm]{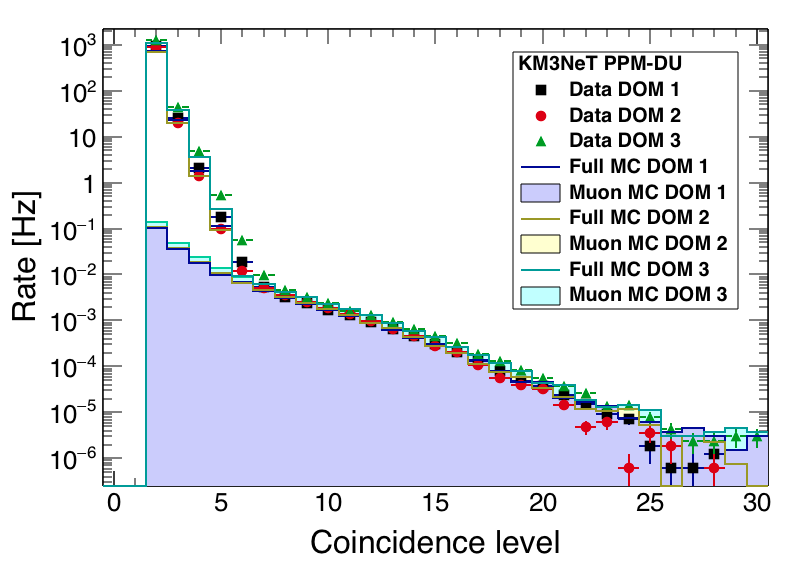}
\caption{Rates of multifold coincidences in a time window of $\unit[25]{ns}$  for the 3 DOMs, compared to the expected   Monte Carlo rates. 
Symbols refer to data, histograms to Monte Carlo simulations. No normalisation factor is applied to Monte Carlo rates.}
\label{f:ratesMultiplicity}
\end{figure}

In Fig.~\ref{f:ratesMultiplicity}   the rates of multifold coincidences in the single DOMs are  shown and  compared to the rates predicted by the Monte Carlo simulation.   
The full Monte Carlo histograms reported in Fig.~\ref{f:ratesMultiplicity} refer to the sum of atmospheric muon events and  $^{40}$K only events.
No normalisation factor is applied to the Monte Carlo events thus showing an excellent absolute agreement between data and Monte Carlo simulations.
At low coincidence multiplicities the signals from $^{40}$K dominate the rates while the muon signature becomes dominant  for coincidence multiplicities exceeding six.

\begin{figure}
\centering
\includegraphics[width=8.5cm]{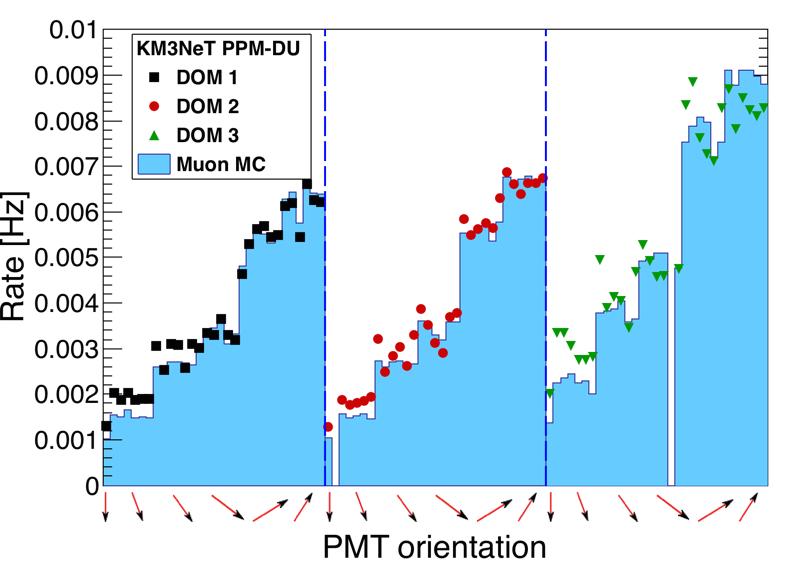}
\caption{Rates of hits on the single DOMs for at least $8\text{-fold}$ coincidences on the respective DOM as a function of the PMT position. 
The data of the three DOMs   are shown in comparison with the atmospheric muon simulation.  No normalisation factor is applied to Monte Carlo rates.}
\label{f:ratesZenith}
\end{figure}

The distribution of hits of the high-multiplicity coincidences ($>7\text{-fold}$) over the different PMTs in each DOM  compared to the corresponding distribution in the muon Monte Carlo simulation is shown in Fig.~\ref{f:ratesZenith}.   
The horizontal axis corresponds to the PMT location in each DOM as mentioned in Sect.~\ref{s:analysis}, starting with the downward looking PMT and subsequently going up the consecutive rings of PMTs.   
The general pattern in all three DOMs clearly shows a higher hit frequency  on the top hemispheres of the DOMs, reflecting the fact that atmospheric muons come from above and demonstrating the directional sensitivity of the DOMs.

\begin{figure}
\centering
{\includegraphics[width=8.5cm]{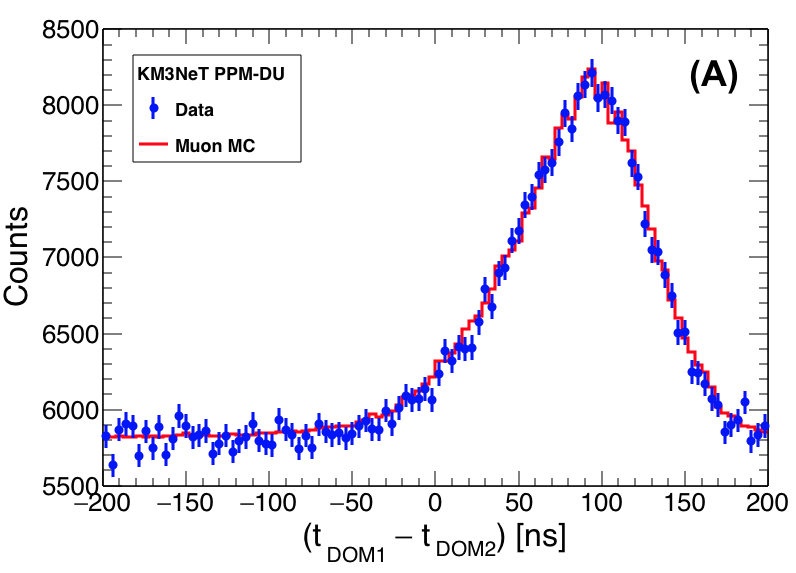}} \\ 
{\includegraphics[width=8.5cm]{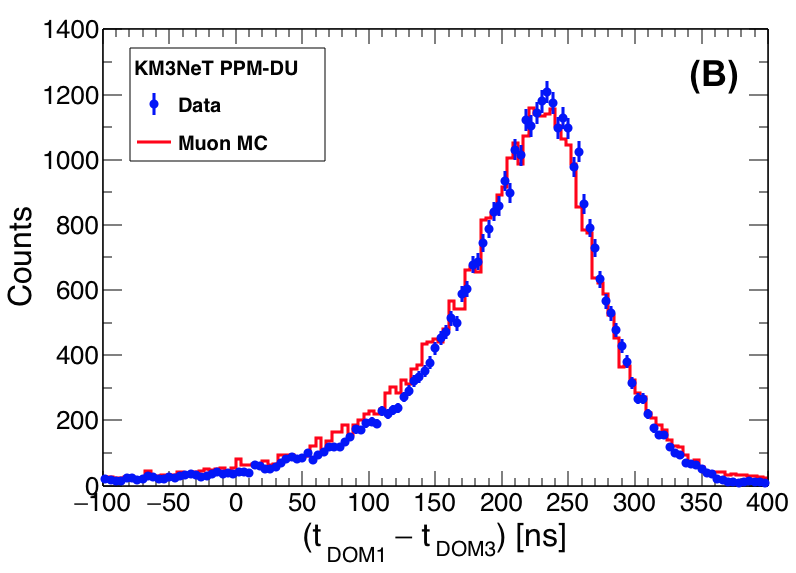}}
\caption{
Time differences between more than twofold coincidences on the different DOMs: (a) DOM~1--DOM~2 and (b) DOM~1--DOM~3 for events when also in DOM~2 a coincidence in time consistent with a muon signal has been detected.
The Monte Carlo distributions are scaled to the total number of events in the data distributions with a factor $\sim10\%$ in order to appreciate the similarity in the shapes.}
\label{f:muondeltaT}
\end{figure}

\subsection{Multi-DOM muon signature}
\label{s:multiDOM}

The distributions of time differences between coincidences on the different DOMs are shown in Fig.~\ref{f:muondeltaT}, together with the predictions from the Monte Carlo muon simulation.  
The time of a coincidence is defined as the start time of the earliest hit in the coincidence.    
The zenith  angles of the incoming muons and their distances to the string determine the spread and overall shape of these distributions. Very good agreement between data and the Monte Carlo simulation is observed. 
The peak due to correlated muon signals in the two  DOMs is   clearly visible on a  flat background due to random uncorrelated signals.  
The additional requirement of a coincidence in the third DOM within a time window consistent with the signal from a muon removes this background, and selects thus an almost background-free sample of muons, as seen in  Fig.~\ref{f:muondeltaT}b.  
The size of this time window is  consistent with the trigger definition  explained in Sect.~\ref{sec:daq}, where a time difference  smaller than $\unit[330]{ns}$ between DOMs triggers the event selection.

Time coincidences between different DOMs  have   been exploited for an accurate time calibration during observing periods that were lacking nanobeacon data.
For this additional time calibration three independent distributions  were used:  
\begin{itemize}
\item time differences between DOM~1 and DOM~2 (no matching coincidence in DOM~3);
\item time differences between DOM~2 and DOM~3 (no matching coincidence in DOM~1);
\item time differences between DOM~1 and DOM~3 when all 3 DOMs are in  coincidence.  
\end{itemize}
The $\chi^2$ difference between the data and MC histograms as given in the formula below was then evaluated as a function of the two independent time offsets of DOM~2 and DOM~3:
$$\chi^2=\sum_i \left[ \sum_j (n_{\text{MC},i,j}-n_{\text{data},i,j})^2 / (n_{\text{MC},i,j}+n_{\text{data},i,j}) \right]$$
Here the summations are over the three distributions  ($i$) and over the nanosecond time bins ($j$). 
The minimum in the resulting $\chi^2$ plane ($x\text{-axis}$: time offset DOM~2, $y\text{-axis}$: time offset DOM~3) was found via a paraboloid fit which provided the corresponding time offsets.
A cross check of this calibration with the nanobeacon calibration demonstrated agreement of the calibrated time offsets within two nanoseconds.

Events with correlated coincidences in all three DOMs were used to reconstruct the zenith angle of the downgoing muons.  
No attempt to reconstruct the azimuth angle of the muons was made. 
The track of a muon can   be parametrised by a time offset $t_0$, the zenith angle $\theta$, the closest distance $d_0$ to the string  and the $z$-position of the closest distance to the string, $z_0$.  
The expected time of arrival of a signal at a DOM in this parametrisation is given by:  
$$t=\frac{ (z-z_0) \cos\theta+\sqrt{n^2-1} \cdot \sqrt{d_0^2+(z-z_0)^2 \sin^2\theta} }{c}+t_0$$ 
where $z$ is the  projection  of the particle on the vertical axis, $n$ is the refractive index of sea water and $c$ is the speed of light.

\begin{figure}
\centering
{\includegraphics[width=8.5cm]{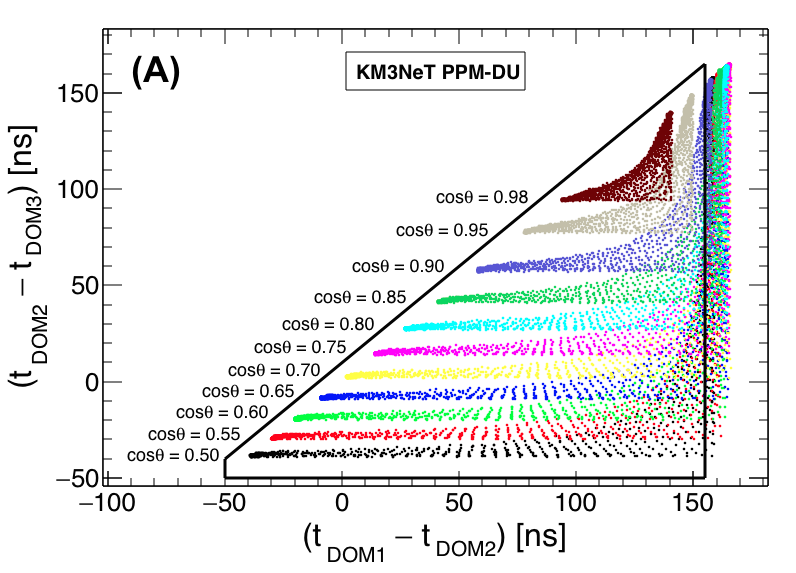}} \\ 
{\includegraphics[width=8.5cm]{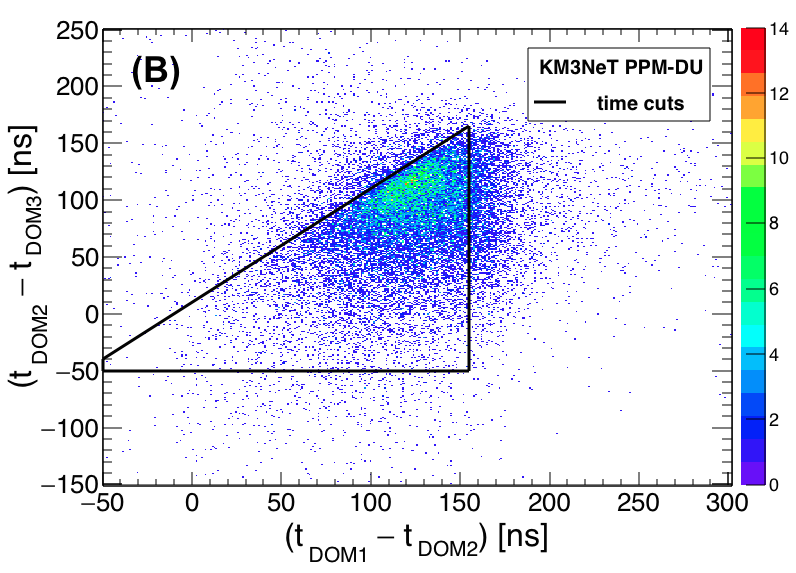}}
\caption{Time differences of the signals between the upper and the middle DOM pair versus the time differences between the signals in the middle and the bottom DOM pair. Every point represents a muon event generated with MC simulations, the zenith angles indicate the Monte Carlo values.  (a) Phase space covered by tracks with different zenith angles for $d_0<\unit[10]{m}$. (b) Phase space covered by the reconstructed time differences before and after selection.}
\label{f:dt1dt2}
\end{figure}

Using only the information of the time differences between the two DOM pairs   cause degeneracies in the track reconstruction, affecting the accuracy of the results.    
In order to reduce the degeneracies, 
only events with $\unit[-50]{ns} < \Delta T_{12} < \unit[155]{ns}$, $\unit[-50]{ns} < \Delta T_{23} < \unit[165]{ns}$, $(\Delta T_{23} - \Delta T_{12}) < \unit[10]{ns}$ were kept. 
The time differences in the signals of the upper and middle DOMs versus the time differences in the middle and lower DOM pair are shown in Fig.~\ref{f:dt1dt2}a for muon Monte Carlo events.
The colours indicate different directions of the muon zenith angle; various  distances of the muon tracks to the detector are covered  in  the vicinity of the detector.   
In  Fig.~\ref{f:dt1dt2}b, the distribution of the reconstructed time differences in the Monte Carlo simulated events before and after selection are shown.    
The selection keeps $\sim67\%$ of all events with coincident signals in all 3 DOMs.

The times of the coincidences in the DOMs were compared to the expected signal times from a possible muon track.   
A $\chi^2$ minimisation scan was performed in flat $\cos\theta$ between 0.5 and 1 (100 steps), corresponding to the assumption of downgoing muons.  
For each of these, the remaining three  parameters ($d_0$, $t_0$, $z_0$) were varied, and the values resulting in the lowest $\chi^2$ were chosen as the final parameters.
Events with  $d_0 > \unit[10]{m}$ are rejected to ensure a good quality of the reconstructed data sample.

\begin{figure}
\centering
{\includegraphics[width=8.5cm]{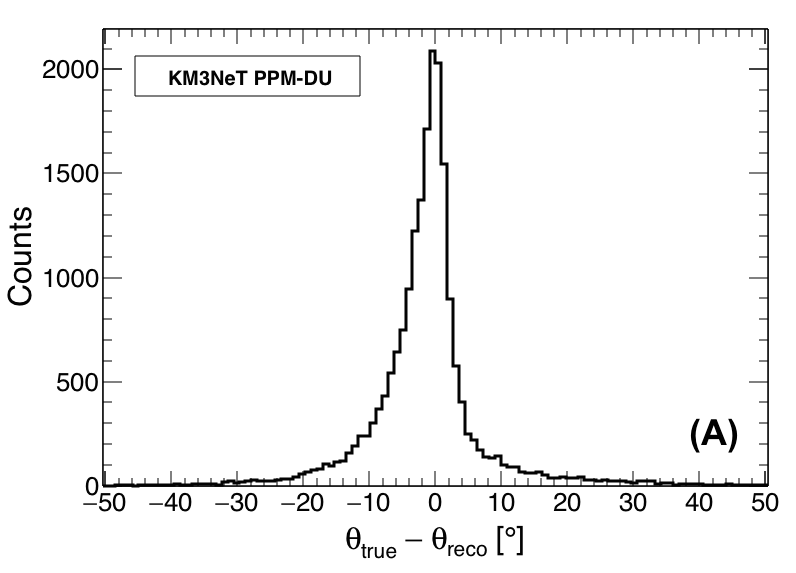}} \\ 
{\includegraphics[width=8.5cm]{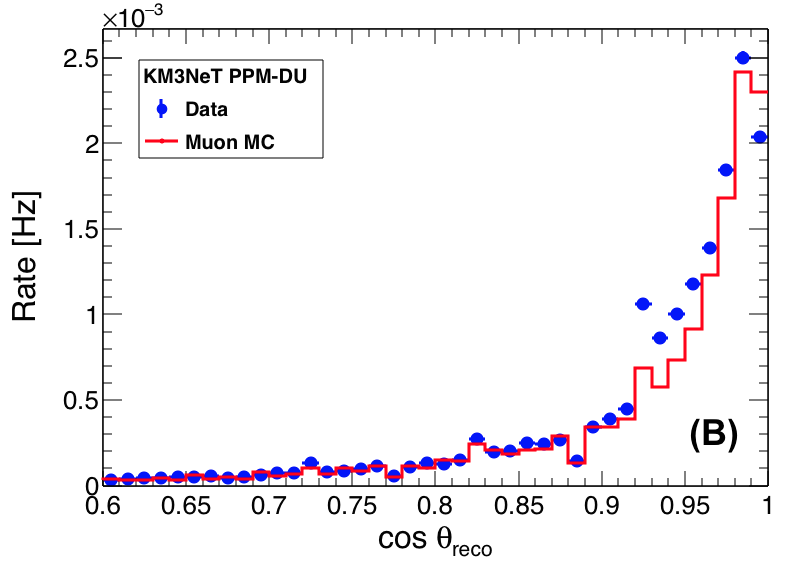}}
\caption{(a) Zenith angular resolution of the tracking algorithm from Monte Carlo simulations. (b) Rates of the reconstructed $\cos\theta$ in   data and   simulation.}
\label{f:zenith_rec_3doms}
\end{figure}

The distribution of the differences between the reconstructed and simulated zenith angle $\theta_{\text{rec}}-\theta_{\text{true}}$ is shown in    Fig.~\ref{f:zenith_rec_3doms}a, resulting in a peak with a FWHM of 7.6 degrees.     
The rates of the reconstructed  $\cos\theta$ for the selected events is shown for both data and Monte Carlo in Fig.~\ref{f:zenith_rec_3doms}b, demonstrating excellent agreement.

\section{Conclusions} 
\label{s:conclusion}

A   prototype of the KM3NeT detection unit was deployed at $\unit[3500]{m}$ in the Mediterranean Sea in May  2014 and was operated until its decommissioning in July 2015. 
The complete marine operations chain for the installation of the DU (DU deployment on the sea bed, submarine connections with ROV, unfurling procedures) was fully  successful. 
This prototype project validated the DU structure at the depth of $\unit[3500]{m}$  providing a test bench for the   operation and data handling tools.  
The prototype was also a tool for testing the software architecture developed for the full scale KM3NeT detector.  
The prototype allowed for long-term monitoring of the optical background ($^{40}$K decay and bioluminescence), improving our knowledge of the marine site.

The procedure for time calibration exploiting $^{40}$K decays and  LED beacons was demonstrated successfully with nanosecond stability.  
The timing information of the signals was  exploited to identify  correlated signals from atmospheric muons in the DOMs.  
Excellent agreement was found in the expected time distributions of signals from muons with simulated signal distributions.   
With the three DOMs, a high purity sample of atmospheric muons was isolated and excellent agreement was found between the observed and simulated   distributions.  
The success of this prototype project paves the path to the forthcoming installation of detection units at the Capo Passero and Toulon sites.  


\bibliographystyle{elsarticle-num}
\bibliography{}

\end{document}